\documentstyle[psfig]{l-aa}

\begin{document}

\thesaurus{ 06(08.02.6; 08.09.2 (CoD-33 7795); 08.12.1; 08.16.5) }

\title{Spectrum and proper motion of a brown dwarf companion 
of the T Tauri star CoD$-33 ^{\circ} 7795$\thanks{Based on observations
obtained at the European Southern Observatory, Cerro Paranal, 
partly from program 65.L-0144 and partly based on public data released 
from FORS2 technical observations at the VLT Kueyen telescope} }

\author{ R. Neuh\"auser\inst{1} \and E.W. Guenther\inst{2} \and M.G. Petr\inst{3}
\and W. Brandner\inst{4} \and N. Hu\'elamo\inst{1} \and J. Alves\inst{5} }

\offprints{R. Neuh\"auser, rne@mpe.mpg.de }

\institute{MPI f\"ur extraterrestrische Physik, D-85740 Garching, Germany
\and Th\"uringer Landessternwarte Tautenburg, Sternwarte 5, D-07778 Tautenburg, Germany
\and MPI f\"ur Radioastronomie, Auf dem H\"ugel 69, D-53121 Bonn, Germany
\and University of Hawaii, Institute for Astronomy, 2680 Woodlawn Dr., Honolulu, HI 96822, USA
\and European Southern Observatory, Karl-Schwarzschild-Stra\ss e 2, D-85748 Garching, Germany
}

\date {Received 26 June 2000; 15 July 2000}

\maketitle

\markboth{Neuh\"auser et al.: Brown dwarf companion of CoD$-33 ^{\circ} 7795$}{}

\begin{abstract}

We present optical and infrared spectra as well as the proper motion 
of an H=12 mag object $2^{\prime \prime}$ off the $\sim 5$ mag brighter 
spectroscopic binary star CoD$-33 ^{\circ} 7795$ (=TWA-5), a member of the 
TW Hya association of T Tauri stars at $\sim 55$ pc. It was suggested 
as companion candidate by Lowrance et al. (1999) and Webb et al. 
(1999), but neither a spectrum nor the proper motion of the
faint object were available before. Our spectra taken with 
FORS2 and ISAAC at the ESO-VLT reveal that the companion 
candidate has spectral type M8.5 to M9.
It shows strong H$\alpha$ emission and weak Na I absorption,
both indicative of a young age. The faint object is clearly detected 
and resolved in our optical and infrared images, 
with a FWHM of $0.18 ^{\prime \prime}$ in the FORS2 image.
The faint object's proper motion, based on two year epoch difference,
is consistent with the proper motion of CoD$-33 ^{\circ} 7795$
by 5 Gaussian $\sigma$ significance. From three different 
theoretical pre-main sequence models, we estimate the
companion mass to be between $\sim 15$ and 40~M$_{\rm jup}$,
assuming the distance and age of the primary.
A slight offset between the VLT and HST images with an epoch 
difference of two years can be interpreted as orbital motion.
The probability for chance alignment of such a late-type object
that close to CoD$-33 ^{\circ} 7795$ with the correct proper motion 
is below $7 \cdot 10^{-9}$. Hence, the faint object is 
physically associated with CoD$-33 ^{\circ} 7795$, the 4th brown 
dwarf companion around a normal star confirmed by both spectrum and 
proper motion, the first around a pre-main sequence star.


\end{abstract}

\section{Introduction: Brown dwarfs as companions}

Despite extensive imaging surveys (e.g. Oppenheimer et al. 2000), only three
brown dwarfs were confirmed so far by both spectroscopy and proper motion
as companions to normal stars: Gl~229~B (Nakajima et al. 1995, 
Oppenheimer et al. 1995), G~196-3~B (Rebolo et al. 1998), and
Gl~570~D (Burgasser et al. 2000).
A few more candidates were presented, GG~Tau~Bb (White et al. 1999),
CoD$-33 ^{\circ} 7795$~B (Lowrance et al. 1999, henceforth L99; 
Webb et al. 1999, W99), and HR~7329~B (Lowrance et al. 2000), 
but either spectroscopy or proper motions were not available. 
Brown dwarfs and L-dwarfs can also have companions 
(Basri \& Mart\'\i n 1999, Mart\'\i n et al. 1999).
Radial velocity surveys yielded a large number of 
planet candidates, but only few brown dwarfs are 
among them, e.g. HD 10697 (Zucker \& Mazeh 2000).

Because young objects are still relatively luminous
due to ongoing accretion and/or contraction
(Burrows et al. 1997, Brandner et al. 1997, Malkov et al. 1998),
imaging surveys for sub-stellar objects in star forming regions
or as companions to isolated young nearby stars should be more fruitful.
E.g., L99 and W99 found a faint object 
called CoD$-33 ^{\circ} 7795$~B just $2 ^{\prime \prime}$ north of the 
isolated M1.5-type T Tauri star CoD$-33 ^{\circ} 7795$~A
(Gregorio-Hetem et al. 1992), a kinematic member of the nearby
TW Hya association (TWA, see Kastner et al. 1997). 
The Hipparcos satellite obtained the parallaxes of four out of 14
TWA members, so that we can assume the mean distance of those 
four stars for the other stars not observed by Hipparcos 
(including CoD$-33 ^{\circ} 7795$~A), namely $55 \pm 16$ pc.

The companion candidate CoD$-33 ^{\circ} 7795$~B is $\sim 5$ mag 
fainter than the primary star in the infrared, and its IHJK colors 
are consistent with spectral type M8 to M8.5 (L99, W99).
Based on its colors, its small separation from CoD$-33 ^{\circ} 7795$~A,
and its galactic latitude, it was concluded that this object could
well be a brown dwarf companion, but neither a spectrum nor the
proper motion were available for corroboration.
Weintraub et al. (2000) presented additional HST NICMOS narrow-band
filter photometry, also consistent with a young late M-type brown dwarf,
but the epoch difference (0.2 yrs) between their and previous images were not
sufficient for obtaining the proper motion of CoD$-33 ^{\circ} 7795$~B.

\section{The spectral type of CoD$-33 ^{\circ} 7795$~B}

An optical spectrum of CoD$-33 ^{\circ} 7795$~B was obtained with the 
FOcal Reducer/low dispersion Spectrograph 2 (FORS2) at the European Southern 
Observatory (ESO) 8.2m telescope Kueyen, Unit Telescope 2 (UT2) of the 
Very Large Telescope (VLT).
The 30 min exposure spectrum in the 6000 to 9000 \AA~range (R=680) using 
grism 300I and order separation filter OG590 was taken during a technical 
night on 23 Feb 2000. The $0.7 ^{\prime \prime}$ slit was positioned 
just on object B in E-W direction.

Standard data reduction was done with MIDAS.
The final spectrum is shown in figure 1.
The spectral type of CoD$-33 ^{\circ} 7795$~B 
is M8.5 to M9 according to different spectral indices
(see also Kirkpatrick et al. 1991).
The equivalent width of the H$\alpha$ emission is $\sim 20$\AA ,
stronger than in old M8-M9 dwarfs.
The Na I doublet line at 8183 and 8195\AA~is slightly weaker than
in the standards, which is indicative of low gravity (Kirkpatrick et al. 
1991). Both the strong H$\alpha$ emission and the weak Na absorption
indicate a young age. The spectral resolution is too low to split
the Na I doublet or to resolve the Li 6708\AA~line
(next to the TiO 6713\AA~and Ca 6718\AA~lines).

\begin{figure}
\vbox{\psfig{figure=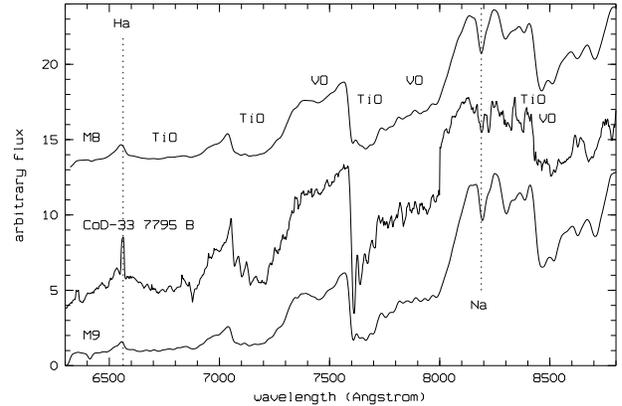,width=9.5cm,height=6.5cm,angle=270}}
\caption{Our optical spectrum of CoD$-33 ^{\circ} 7795$~B compared with
M8 dwarfs (average of LHS 2243, LHS 2397 A, LP 412-31, and vB 10) and 
M9 dwarfs (average of BRI 1222-1222, LHS 2065, LHS 2924, and TVLM 868-110639),
showing that our object is M8.5 to M9 (comparison spectra from K. Luhman).
Strong H$\alpha$ and weak Na indicate a young age}
\end{figure}

An H-band spectrum (R$\simeq 500$) was obtained on 16 Apr 2000 with the Infrared 
Spectrograph and Array Camera (ISAAC) at the ESO 8.2m telescope Antu (VLT-UT1).
The spectrum consists of 20 co-added 60s
exposures through a $0.6 ^{\prime \prime}$ slit, aligned 
neither along the position angle of the pair nor perpendicular to it, 
but in between those two positions, so that the two objects are well 
separated and that the flux from the companion 
candidate is several times larger than the flux from the bright star. 

\begin{figure}
\vbox{\psfig{figure=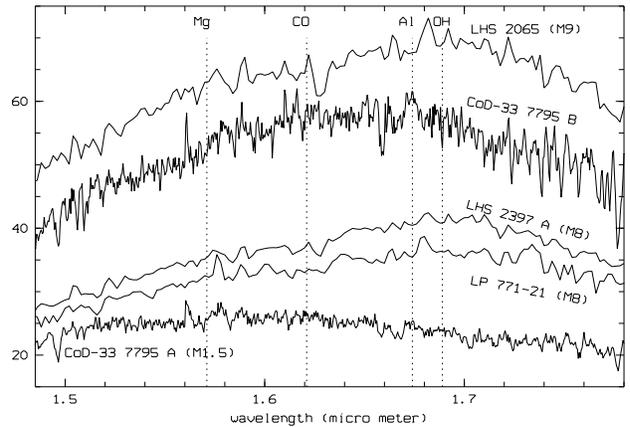,width=9.5cm,height=6.5cm,angle=270}}
\caption{Our H-band spectrum of CoD$-33 ^{\circ} 7795$~B compared with
the M8 dwarfs LHS 2397 A and LP 771-21 as well as the M9 dwarf LHS 2065
(Delfosse et al. 1998) showing that our object is M8.5 to M9.
Also plotted is CoD$-33 ^{\circ} 7795$~A}
\end{figure}

After standard data reduction, we modelled and subtracted the flux 
of the bright star from the faint object's spectrum at each wavelength.
Mainly due to the slope of the continuum (figure 2), 
CoD$-33 ^{\circ} 7795$~B has spectral type M8.5 to M9,
consistent with the IHJK colors (L99 and W99).
Recently, Schneider et al. (2000) presented an HST/STIS spectrum of 
CoD$-33 ^{\circ} 7795$~B, which is in good agreement with our results.

\section{The proper motion of CoD$-33 ^{\circ} 7795$~B}

CoD$-33 ^{\circ} 7795$~B was detected by L99 using HST NICMOS on 
25 Apr 1998 in the F160W filter, located $0.04 \pm 0.01 ^{\prime \prime}$
west\footnote{As noticed by Weintraub et al. (2000),
there is a sign error in the right ascension offset in both L99 and W99.}
and $1.95 \pm 0.01 ^{\prime \prime}$ north of CoD$-33 ^{\circ} 7795$~A,
corresponding to a separation of $\rho = 1.96  \pm 0.01 ^{\prime \prime}$
and a position angle of $\theta = -1.2 \pm 0.1 ^{\circ}$.
On 12 Jul 1998, Weintraub et al. (2000) detected the faint object,
also using HST NICMOS, but with narrow band filters, 
located $0.038 \pm 0.001 ^{\prime \prime}$ west 
and $1.960 \pm 0.006 ^{\prime \prime}$ north of the bright star,
corresponding to $\rho = 1.960 \pm 0.006 ^{\prime \prime}$
and $\theta = -1.11 \pm 0.03 ^{\circ}$.
The precision in Weintraub et al. (2000) is higher than in L99,
because the latter used the coronograph that makes it difficult
to determine the centroid.

\begin{figure*}[t]
\vbox{\psfig{figure=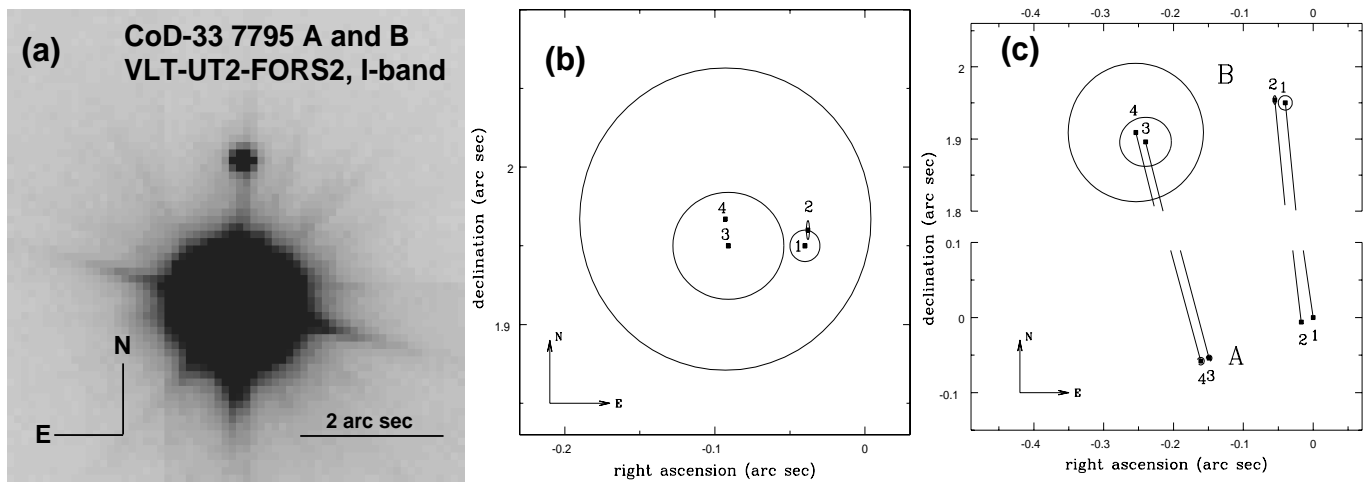,width=18cm,height=6.5cm,angle=270}}
\caption{(a) FORS2 acquisition image of CoD$-33 ^{\circ} 7795$~A and B,
where star A is saturated, FWHM of object B is $0.18 ^{\prime \prime}$.
(b) and (c) Position of companion candidate B relative to star A,
plotted are $\alpha$ and $\delta$ offset as given in the text.
Data points 1 \& 2 are from NICMOS (1998.3 and 1998.5), 
point 3 from FORS2 (2000.1), and point 4 from ISAAC (2000.3), 
all with $1 \sigma$ error ellipses.
(b) Relative location of object B: Star A is always at $(\alpha ,\delta)=(0,0)$.
If the error ellipses would be disjunct by more than to be allowed
for orbital motion ($13.4 \pm 4.2$ mas/yr, see text), 
object B would be unrelated.
The objects do not have significantly different relative motion.
(c) With proper motion: Star A starts at $(\alpha ,\delta)=(0,0)$
and then moves to the south-west. 
Here, if the error ellipses would overlap, object B would be unrelated.
The small $1.2 \sigma$ offset of object B in the FORS2 image relative
to the HST images is consistent with orbital motion (see text).
Because object B is clearly co-moving with star A, it is a companion}
\end{figure*}

We present two new images of CoD$-33 ^{\circ} 7795$~B:
A 1s exposure FORS2 I-band image taken during a technical night 
on 21 Feb 2000 with the high resolution collimator 
($0.1 ^{\prime \prime}$/pixel) and a 2s exposure ISAAC 
acquisition image ($0.147 ^{\prime \prime}$/pixel) taken on 16 Apr 2000 through 
a narrow band filter centered on $1.64 \mu $m ($\Delta \lambda = 0.025 \mu $m).
In both images, the central pixels of the bright star are saturated,
which makes it difficult to determine the centroid; 
we fitted isophots in the unsaturated part of the PSF.
The FWHM of the faint object on the FORS2 image is only 
$0.18 ^{\prime \prime}$, so that this image may well be the 
sharpest optical image ever taken from the ground (figure 3a).

In the FORS2 image, the companion candidate is located $0.091 \pm 0.037 ^{\prime \prime}$
west and $1.950 \pm 0.034 ^{\prime \prime}$ north of the bright star,
corresponding to $\rho = 1.952 \pm 0.050 ^{\prime \prime}$ and
$\theta = -2.7 \pm 1.2 ^{\circ}$, and in the ISAAC image,
the companion candidate is located $0.093 \pm 0.097 ^{\prime \prime}$
west and $1.967 \pm 0.096 ^{\prime \prime}$ north of the bright star,
corresponding to $\rho = 1.969 \pm 0.091 ^{\prime \prime}$ and
$\theta = -3.2 \pm 3.0 ^{\circ}$.
The errors include uncertainties in the north-south alignment.

In figure 3b, we plot the four positions of the companion candidate B
with respect to star A with their error ellipses. 
If the error ellipses are disjunct by more than expected for
orbital motion ($13.4 \pm 4.2$ mas/yr, see below), 
object B could not be a co-moving companion.
If the error ellipses do overlap, this does not prove object B
to be a companion. Whether we can already show that the
motion of CoD$-33 ^{\circ} 7795$~B relative to A is inconsistent with 
B being an unrelated field star, depends on the proper motion of star A.
The proper motion was published by W99. 
In the Tycho catalog (H{\o}g et al. 2000), we found
$\mu _{\alpha} = -81.6 \pm 2.5$ and $\mu _{\delta} = -29.4 \pm 2.4$ mas/yr.

In figure 3c, we plot star A first on 25 Apr 1998 at $(\alpha ,\delta)=(0,0)$,
then on 12 Jul 1998 south-west of it as given by its proper motion,
and then on 21 Feb and 16 Apr 2000 even more south-west; the errors in
the 2nd to 4th epoch locations are given by the error of the
proper motion. In addition, we plot the offset of object B relative to 
star A with errors given by the errors of the measured offsets 
and the proper motion of star A.
Object B is clearly co-moving with star A. If object B would 
be an unrelated field object, it should not be co-moving with A, but either
be a non-moving background object or a foreground object with different motion
(different parallactic motion would be negligible, even if unrelated, because the 
epoch difference between the HST and VLT images is close to an integer number of years). 
The error ellipses do not overlap. The proper motions of A and B are similar, 
namely by $2 \sigma$ regarding their amount and by $3 \sigma$ 
regarding their direction. Hence, we have in total a $5 \sigma$ significance 
for the pair being a common proper motion pair.

\section{The mass of CoD$-33 ^{\circ} 7795$~B}

Based on its spectral type and magnitude (H$=12.14 \pm 0.06$ mag, L99),
CoD$-33 ^{\circ} 7795$~B would be located at $\sim 18.5$ pc, if it would
be main-sequence dwarf (M$_{\rm H} = 10.8$ mag, Kirkpatrick \& McCarthy 1994).
From the six objects with M$_{\rm H} \ge 10$ mag found within 5 pc
around the Sun, we can estimate the probability for
chance alignment of CoD$-33 ^{\circ} 7795$~B within 
$1.96 ^{\prime \prime}$ around star A to be $7 \cdot 10^{-9}$.
Given the very sparse space density of T Tauri stars in the TWA
region, the probability for CoD$-33 ^{\circ} 7795$~B to be a
free-floating young TWA brown dwarf, unrelated to star A, 
is of the same order. Thus, there is a high probability that 
component B is a physical companion to star A.

CoD$-33 ^{\circ} 7795$~A is a spectroscopic binary (W99).
For an equal-mass binary at $\sim 55$ pc, the age is $12 \pm 6$ Myrs
(Weintraub et al. 2000). We can assume the same age for its companion.
Hence, for its young age and spectral type,
CoD$-33 ^{\circ} 7795$~B is below the sub-stellar limit
according to different sets of tracks and isochrones
(e.g. Baraffe et al. 1998). Hence, it is a brown dwarf.

The mass of each component in the spectroscopic binary CoD$-33 ^{\circ} 7795$~A,
assuming that both components have equal masses, is $0.75 \pm 0.15$~M$_{\odot}$
(Weintraub et al. 2000 using Baraffe et al. 1998 tracks).
Thus, the separation $1.96 \pm 0.01 ^{\prime \prime}$ at $55 \pm 16$ pc distance
corresponds to a projected separation of $108 \pm 16$~AU 
and to an orbital period of $916 \pm 301$~yrs.
Assuming a circular orbit viewed pole-on, we expect
$13.4 \pm 4.2$ mas/yr orbital motion.

The location of object B relative to star A in the FORS2 image 
is $\sim 1 \sigma$ deviant from the HST images (figure 3b):
object B lies $0.054 \pm 0.038 ^{\prime \prime}$ west of star A.
This can be interpreted as first indication for orbital motion
after the two year epoch difference.
The alternative interpretation that object B is a fast moving
foreground star, is extremly unlikely (see above).
If this slight deviation indeed is orbital motion, the inclination
is not edge-on, because we see motion in the plane of the sky.
Given the good seeing and image quality at the VLT, the errors in 
the location of object B relative to star A should improve, if one 
can obtain unsaturated images. Then, one can detect 
curvature in the orbit within a few years.

Given the young age and spectral type M8.5 to M9 of CoD$-33 ^{\circ} 7795$~B, 
its effective temperature -- using the scale intermediate between giant 
and dwarfs provided by Luhman (1999) -- is $2550 \pm 150$~K, where the error 
comes from the error in the Luhman scale and the spectral type ($\pm 0.25$ sub-types).
This results in a bolometric luminosity of 
$\log$ (L$_{\rm bol}/$L$_{\odot}) = -2.60 \pm 0.29$ 
(using B.C.$_{\rm H}=2.8$ mag).

Comparing these numbers with theoretical models, we can estimate its mass:
From the Burrows et al. (1997) models, we obtain $\sim 30$~M$_{\rm jup}$.
According to Baraffe et al. (1998), the object is located on the 10 Myrs 
isochrone (co-eval with the primary) with a mass of $30 \pm 10$~M$_{\rm jup}$.
With the new Chabrier et al. (2000) models, the companion 
has a mass of $20 ^{+10} _{-5}$~M$_{\rm jup}$ for an age of 1 to 20 Myrs.
Overall, a range of $\sim 15$ to 40~M$_{\rm jup}$ is reasonable.
All those models, however, are uncertain at the young age of our object.

Because CoD$-33 ^{\circ} 7795$~A is a spectroscopic binary
and because it may soon be possible to detect orbital motion of the
companion brown dwarf, masses and/or mass ratios might be determined soon.
Finally, all three objects should be co-eval, so that this triple system
will be a good test case for theoretical evolutionary tracks and isochrones.

\acknowledgements{
We would like to thank F. Comer\'on for his support from ESO, 
J.-G. Cuby and C. Lidman for the ISAAC service mode observations,
X. Delfosse and K. Luhman for providing their spectra 
of late-type dwarfs in electronic form (partly via F. Comer\'on), 
as well as G. Wuchterl and S. Frink for very usefull discussion.}

{}

\end{document}